\documentclass[CEJP,PDF,DVI]{cej} % use PDF command to enable PDFLaTeX driver
\usepackage{layout}
\usepackage{amsmath}
\usepackage{textcomp}

\usepackage{graphicx}

\usepackage{dcolumn}
\usepackage{bm}

\title{Flowing Liquid Crystal Simulating the Schwarzschild Metric}

\author{Erms R. Pereira and Fernando Moraes$^\dagger$}

\institute{Departamento de F\'{\i}sica, CCEN,  Universidade Federal 
da Para\'{\i}ba, Cidade Universit\'{a}ria, 58051-900 Jo\~{a}o Pessoa, PB,
Brazil}

\abstract{
We show how to simulate the equatorial section of the Schwarzschild metric through a flowing liquid crystal in its nematic phase. Inside a liquid crystal in the nematic phase, a traveling light ray feels an effective metric, whose properties  are linked to perpendicular and parallel refractive indexes, $n_o$ e $n_e$ respectively, of the rod-like molecule of the liquid crystal. As these indexes depend on the scalar order parameter of the liquid crystal, the Beris-Edwards hydrodynamic theory is used to connect the order parameter with the velocity of a liquid crystal flow at each point. This way we  calculate a radial velocity profile that simulates the equatorial section of the Schwarzschild metric, in the region outside of Schwarzschild's radius,  in the nematic phase of the liquid crystal. In our model, the higher flow velocity can be on the order of some meters per second.}

\keywords{Analogue Models, Black Hole Geometry, Liquid Crystal}
\pacs{04.70.Bw, 42.70.Df, 47.57.Lj, 61.30.Dk}

\begin{document}
\maketitle

$^\dagger$E-mail: moraes@fisica.ufpb.br

%%%%%%%%%%%%%%%%%%%%%%%%%%%%%%%%%%%%%%%%%%%%%%%%%%%%%%%%%%%%%%%%%%%%%%%%%%%%%%
\section{Introduction}
%%%%%%%%%%%%%%%%%%%%%%%%%%%%%%%%%%%%%%%%%%%%%%%%%%%%%%%%%%%%%%%%%%%%%%%%%%%%%%
Research in analogue models explores the similarities among  areas of  physics as different among themselves as general relativity and condensed matter \cite{bjp}. It  maps the main characteristics from a subject to another  bringing in phenomenological insights even if  there is not  an obvious connection between them. For example,  defects in crystals can be described by an effective Riemann-Cartan spacetime \cite{kleinert1,kleinert2}. In fluid systems, Unruh's effect \cite{unruh} shows the similarity between the phonon creation in a radial flowing fluid and Hawking's effect around a black hole. Also,  many analogue models for different black hole's aspects have appeared in the literature \cite{novello}.  These procedures admit, not  only new understanding on the subjects involved, but the possibility to make well controlled experiments with  systems , like dielectric fluids or liquid crystals, as well. In this paper we are particularly interested in these latter ones.

Liquid crystals are materials composed by anisotropic molecules, like rods, having different phases between the liquid and crystalline one. In one of these, the nematic phase,  the rod-like molecules  on the average point towards a specific direction, represented by the vector $\vec{n}$ \cite{gennes,kleman}. As shown by Kline and Kay \cite{kline}, Fermat's principle can be used to calculate light's trajectories in inhomogeneous anisotropic media like liquid-crystalline materials. This says that, among all the possible paths linking the points $A$ and $B$, the light will choose the one that minimizes the integral $\int_{A}^{B}N(\vec{r})dl$, where $N(\vec{r})$ is the refractive index of light at the position indicated by the vector $\vec{r}$ and $l$ is a parameter with the dimension of length. This computation of light's path can be substituted by a geometric approach \cite{satiro} and it is equivalent to say that the light, within a liquid crystal in the nematic phase, does not feel the geometric properties of the laboratory, but feels the properties of an effective space, of an effective metric. In this geometric point of view, the light trajectories are geodesics in the effective geometry. This makes a connection between nematic liquid crystals and space-time. For example, hedgehog-like defects in the nematic phase simulate the  global monopole space-time \cite{satiro}, while the disclination-like defects simulate the metric of a cosmic string \cite{satiro,carvalho}.

The geometric properties of the ``space-times" corresponding to the defects in liquid crystals depend on the perpendicular and parallel refractive indexes of the liquid crystal molecule in use, $n_o$ e $n_e$ respectively. As these indexes are temperature dependent, it's possible to describe them in terms of a scalar function that measures how ordered is the nematic phase: the scalar order parameter $q$. If, for different points of the liquid crystalline material, there are different values of local velocity, then the Beris-Edwards hydrodynamic theory \cite{beris} can relate the local velocity of the liquid crystalline bulk to the order parameter $q$. This way,  different velocity profiles generate different refractive index dependent functions and, thus, new effective metrics.

With these ideas in mind, we study the case where light is propagating in the plane $z=cte.$ of a configuration like a $(k=1,c=0)$ disclination defect. This defect is a cylindrically symmetric linear one generated as follows. We assume that the liquid crystal flows radially inwards on a flat surface with a drain as in a sink. This tend to align the molecules radially in the configuration known as radial disclination with unit topological charge \cite{kleman,satiro}. Far from this configuration, the disordered phase is recovered. When a velocity field is applied in the bulk, the molecules orient according to  the velocity gradient, generating an orientational order and a non-null value of the scalar order parameter $q$, that measures the strength of the ordering.  Therefore, we calculate the velocity profile $v(r)$, that is the velocity of the liquid crystalline bulk at the distance $r$ from the core of the configuration. By doing this we simulate the equatorial section, $\theta=\pi/2$, of the Schwarzschild metric:
\begin{align}
\label{schild}
%\begin{split}
ds^2=-\left(1-\frac{2M}{r}\right)dt^2+\frac{dr^2}{\left(1-\frac{2M}{r}\right)}+r^2\left(d\theta^2+\sin^2\theta d\phi^2\right).
%\end{split}
\end{align}

This work is presented in the following way. First, we show the effective metric that describes the light propagation around a $(k=1,c=0)$ disclination defect and how this propagation can be described by the order parameter $q$ of the liquid crystalline material. Afterwards, we consider the liquid crystal flowing radially and we use the Beris-Edwards theory to analyze the dependence of the order parameter of the material with the flowing velocity module. In these two cases we consider the more general situation of three space dimensions. Finally, we employ the result from the second part in the first and we compare with the Schwarzschild metric written in isotropic coordinates \cite{inverno}.

%%%%%%%%%%%%%%%%%%%%%%%%%%%%%%%%%%%%%%%%%%%%%%%%%%%%%%%%%%%%%%%%%%%%%%%%%%%%%%%%%%%%%%%%%%%%
\section{Geometric Model}
%%%%%%%%%%%%%%%%%%%%%%%%%%%%%%%%%%%%%%%%%%%%%%%%%%%%%%%%%%%%%%%%%%%%%%%%%%%%%%%%%%%%%%%%%%%%
%%%%%%%%%%%%%%%%%%%%%%%%%%%%%%%%%%%%%%%%%%%%%%%%%%%%%%%%%%%%%%%%%%%%%%%%%%%%%%%%%%%%%%%%%%%%

%%%%%%%%%%%%%%%%%%%%%%%%%%%%%%%%%%%%%%%%%%%%%%%%%%%%%%%%%%%%%%%%%%%%%%%%%%%%%%%%%%%%%%%%%%%%

The analogy between  Fermat's principle and the calculus of null geodesics in a Riemannian manifold has been known for a long time \cite{born}. This relationship can be represented \cite{satiro} by the following equation when we treat the extraordinary light \cite{nota} paths :
\begin{equation*}
	N_e^2dl^2=\sum_{i,j}g_{i,j}dx^idx^j,
\end{equation*}
being $N_e(\vec{r})$ the refractive index of the extraordinary light. In the $(k=1,c=0)$ disclination case, the effective metric is \cite{satiro}
\begin{equation}
\label{disclii}
	ds^2=b^2d\tilde{r}^2+\tilde{r}^2d\phi^2+dz^2,
\end{equation}
where $b=\frac{n_o}{n_e}$, $\tilde{r}\equiv n_er$ and remembering that $n_o$ and $n_e$ are respectively the refractive indexes of the liquid crystal molecule.

Theses indexes can be written as temperature functions by \cite{li}
\begin{equation}
\label{indices}
	\begin{split}
		&n_o\left(T\right)=A-BT-\frac{\left(\Delta n\right)_0}{3}\left(1-\frac{T}{T_c}\right)^{\beta},\\
		&n_e\left(T\right)=A-BT+\frac{2\left(\Delta n\right)_0}{3}\left(1-\frac{T}{T_c}\right)^{\beta}.
	\end{split}	
\end{equation}
Here, $A$, $B$, $\beta$, $\left(\Delta n\right)_0$ and $T_c$ are constants that characterize the material, where $\left(\Delta n\right)_0$ is the birefringence of the liquid crystal in the crystalline phase and $T_c$ is the transition temperature between the nematic and isotropic phases. Considering the thermal dependence of the order parameter $q$ through Haller's approximation \cite{haller}
\begin{equation}
	q\left(T\right)=\left(1-\frac{T}{T_c}\right)^{\beta}\nonumber,
\end{equation}
the equation (\ref{indices}) changes to %a equao  modifica-se para
\begin{equation}
\label{ind}
	\begin{split}
		&n_o\left(q\right)=A-BT_c\left(1-q^{1/\beta}\right)-\frac{\left(\Delta n\right)_0q}{3},\\
		&n_e\left(q\right)=A-BT_c\left(1-q^{1/\beta}\right)+\frac{2\left(\Delta n\right)_0q}{3}\nonumber,
	\end{split}	
\end{equation}
and thus the equation (\ref{disclii}) becomes

\begin{equation*}
%\begin{split}
	ds^2=\left[\frac{A-BT_c\left(1-q^{1/\beta}\right)-\frac{\left(\Delta n\right)_0q}{3}}{A-BT_c\left(1-q^{1/\beta}\right)+\frac{2\left(\Delta n\right)_0q}{3}}\right]^2d\tilde{r}^2+\tilde{r}^2d\phi^2+dz^2.
%\end{split}
\end{equation*}

The theory developed in \cite{satiro} does not take into account the time coordinate but this coordinate must be taken into account in further calculations. In fact, the time coordinate  enters the equation above in the simplest way, since the geodesic equations of a manifold with the metric $ds^2=A(r)dr^2+r^2\left[B(r)d\theta^2+C(r)\sin^2\theta d\phi^2\right]$ are the null geodesics of the manifold with the metric $d\tilde{s}^2=-dt^2+A(r)dr^2+r^2\left[B(r)d\theta^2+C(r)\sin^2\theta d\phi^2\right]$, resulting in
\begin{equation}
\label{discli3}
	\begin{split}
	ds^2&=-dt^2+\left[\frac{A-BT_c\left(1-q^{1/\beta}\right)-\frac{\left(\Delta n\right)_0q}{3}}{A-BT_c\left(1-q^{1/\beta}\right)+\frac{2\left(\Delta n\right)_0q}{3}}\right]^2d\tilde{r}^2\\
	& \mbox{}+\tilde{r}^2d\phi^2+dz^2.
\end{split}
\end{equation}
We will back to equation (\ref{discli3}) later.

%%%%%%%%%%%%%%%%%%%%%%%%%%%%%%%%%%%%%%%%%%%%%%%%%%%%%%%%%%%%%%%%%%%%
%%%%%%%%%%%%%%%%%%%%%%%%%%%%%%%%%%%%%%%%%%%%%%%%%%%%%%%%%%%%%%%%%%%%
\section{The Beris-Edwards Theory}
%%%%%%%%%%%%%%%%%%%%%%%%%%%%%%%%%%%%%%%%%%%%%%%%%%%%%%%%%%%%%%%%%%%%
%%%%%%%%%%%%%%%%%%%%%%%%%%%%%%%%%%%%%%%%%%%%%%%%%%%%%%%%%%%%%%%%%%%%
The relationship between the scalar order parameter $q$ and the velocity $\vec{v}(\vec{r})$ of the liquid crystalline bulk on the position $\vec{r}$ is expressed by the Beris-Edwards equation \cite{beris}
\begin{equation}\label{edwards}
\left(\partial_t+\vec{v}\cdot\nabla\right)\textbf{Q}-\textbf{S}\left(\textbf{W,\textbf{Q}}\right)=\Gamma \textbf{H},
\end{equation}
where $\textbf{Q}$ is the order parameter tensor and $\Gamma$ is a collective rotational diffusion constant. The tensor $\textbf{S}\left(\textbf{W,\textbf{Q}}\right)$ depends on the gradient velocity tensor $W_{\alpha\beta}=\partial_\beta v_\alpha$ as 
\begin{eqnarray}	\textbf{S}\left(\textbf{W,\textbf{Q}}\right)&=&\left(\xi\textbf{D}+\Omega\right)\left(\textbf{Q}+\textbf{I}/3\right)+\left(\textbf{Q}+\textbf{I}/3\right)\left(\xi\textbf{D}-\Omega\right)\nonumber\\
&-&2\xi\left(\textbf{Q}+\textbf{I}/3\right)Tr\left\{\textbf{Q}\textbf{W}\right\},
\end{eqnarray}
where $\xi$ is the aspect ratio of the liquid crystal molecule, $\textbf{D}=(\textbf{W}+\textbf{W}^T)/2$ and $\Omega=(\textbf{W}-\textbf{W}^T)/2$, the symmetric and antisymmetric parts, respectively, of the tensor $\textbf{W}$. 
The molecular field $\textbf{H}$ on the right-hand side of equation (\ref{edwards}) depends on the derivative of the free energy $\mathcal{F}$ by \cite{yeomans,olmsted}
\begin{align}
\label{H}
	\begin{split}							  \textbf{H}&=-\frac{\delta\mathcal{F}}{\delta\textbf{Q}}+(\textbf{I}/3)Tr\left[\frac{\delta\mathcal{F}}{\delta\textbf{Q}}\right]\\	&=-a\textbf{Q}+b\left(\textbf{Q}^2-(\textbf{I}/3)Tr\left[\textbf{Q}^2\right]\right)-hTr\left[\textbf{Q}^2\right]\\
	&+\kappa\nabla^2\textbf{Q}.
	\end{split}
\end{align}

In this section, we apply the Beris-Edwards equation (\ref{edwards}) in the case where the liquid crystal bulk is moving  radially towards a cylinder with radius $r_0\neq0$ (in microscopic scale) and the initial director is $\vec{n}=\hat{r}$, in cylindrical coordinates. This setup has the purpose to mimic the director of the $(k=1,c=0)$ disclination, whose the effective metric we already know. Moreover, as the bulk flows continually inward, it is important that at $r=r_0$ there are sinks that removes the liquid crystal from the bulk, avoiding that the continuity law be violated. Thus, we use  $\vec{n}=\hat{r}$ to describe the order parameter tensor ($Q_{ij}=n_in_j-\frac{1}{3}\delta_{ij}$) with the velocity dependence
\begin{align}
\label{Q}
	\textbf{Q}(\vec{v}(\vec{r}))=q(v(\vec{r}))\left(\begin{array}{ccc}
	\frac{2}{3}&0&0\\
	0&-\frac{1}{3}&0\\
	0&0&-\frac{1}{3}\\
	\end{array}	\right),
\end{align}
where $v= \left|\vec{v}\right|\equiv\left|\vec{v}(\vec{r})\right|$. As shown in \cite{gennes}, $q(v)\propto\left|\nabla \vec{v}(\vec{r})\right|$ and the cylindrical symmetry of our setup says that the  velocity $v$ doesn't depend on the $\phi$ and $z$ coordinates, allowing one to write $q(v(\vec{r}))\equiv q(r)$. Therefore the flow's direction determines just where the maximum and the minimum of $q$ occur. Expressing $\vec{v}(r)=v(r)\hat{r}$, the tensor 
$\textbf{S}\left(\textbf{W,\textbf{Q}}\right)$ becomes
\begin{align}
\label{S}
	\textbf{S}\left(\textbf{W,\textbf{Q}}\right)=2\xi q\left(r\right)\partial_r\left[v(r)\right]\left(1-\frac{2q(r)}{3}\right)\left(\begin{array}{ccc}
	1&0&0\\
	0&0&0\\
	0&0&0\\
	\end{array}	\right).
\end{align}
Recalling that $q(v)\equiv q(r)$ and substituting (\ref{Q}) in (\ref{H}), we obtain the molecular field $\textbf{H}$
\begin{eqnarray}
\label{Hi}
H_{rr}&=&-\frac{2aq(r)}{3}+\frac{2q^2(r)b}{9}-\frac{4hq^3(r)}{9}\nonumber\\
				 & &\mbox{}+\frac{2\kappa r^{-1}\partial_r\left[r\partial_r\left[q(r)\right]\right]}{3},\nonumber\\
	H_{\theta\theta}=H_{zz}&=&\frac{aq(r)}{3}-\frac{q^2(r)b}{9}+\frac{2hq^3(r)}{9}\nonumber\\
				 & &-\mbox{}\frac{\kappa r^{-1}\partial_r\left[r\partial_r\left[q(r)\right]\right]}{3}.
\end{eqnarray}
The remaining terms are null. 

Substituting (\ref{Q}), (\ref{S}) and (\ref{Hi}) in (\ref{edwards}), we get a matrix equation, whose solution rules the relationship between the scalar order parameter and the modulus of the velocity
\begin{eqnarray}
\label{res}
	q(r)=\frac{3}{2+\frac{3C}{v(r)^{3\xi}}},
\end{eqnarray}
where $C$ is a constant defined by boundary conditions.
%%%%%%%%%%%%%%%%%%%%%%%%%%%%%%%%%%%%%%%%%%%%%%%%%%%%%%%%%%%%%%%%%%%%
%%%%%%%%%%%%%%%%%%%%%%%%%%%%%%%%%%%%%%%%%%%%%%%%%%%%%%%%%%%%%%%%%%%%
\section{The Effective Schwarzschild Metric}
%%%%%%%%%%%%%%%%%%%%%%%%%%%%%%%%%%%%%%%%%%%%%%%%%%%%%%%%%%%%%%%%%%%%
%%%%%%%%%%%%%%%%%%%%%%%%%%%%%%%%%%%%%%%%%%%%%%%%%%%%%%%%%%%%%%%%%%%%
When a metric is described by isotropic coordinates, the spatial section of the new metric is flat and all of it is multiplied by a function. Following the steps presented in \cite{inverno}, we rewrite a metric similar to (\ref{discli3}) in the plane $z=const.$, 
\begin{eqnarray}
\label{com}
	ds^2=-dt^2+p^2(\tilde{r})d\tilde{r}^2+\tilde{r}^2d\phi^2,
\end{eqnarray}
in terms of isotropic coordinates. So (\ref{com}) becomes  
\begin{eqnarray}
\label{2}
	ds^2=-dt^2+\lambda^2\left(\tilde{\rho}\right)\left(d\tilde{\rho}^2+\tilde{\rho}^2d\phi^2\right),
\end{eqnarray}
where $\lambda\left(\tilde{\rho}\right)$ is a function to be determined and $\tilde{\rho}$ is a new radial coordinate. Comparing the radial and angular terms from the last two equations  we get
\begin{align}
\label{3}
	\begin{split}
		p(\tilde{r})d\tilde{r}&=\lambda\left(\tilde{\rho}\right)d\tilde{\rho},\\
		\tilde{r}&=\lambda\left(\tilde{\rho}\right)\tilde{\rho}.
	\end{split}
\end{align}
For the metric (\ref{2}) to have null geodesics that \textbf{coincide} with the null geodesics of the Schwarzschild one (\ref{schild}), \textit{i.e.}, for the metric (\ref{2}) to be conformally related or conformal to (\ref{schild}) , we find in \cite{novello,inverno} that
\begin{eqnarray}
\label{4}	\lambda\left(\tilde{\rho}\right)=\frac{\left(1+\frac{M}{2\tilde{\rho}}\right)^3}{\left(1-\frac{M}{2\tilde{\rho}}\right)},
\end{eqnarray}
where $M$, in terms of the liquid crystal, represents the strength of the cylinder, that creates the disclination-like director, to divert the light.
Substituting (\ref{4}) in (\ref{3}) and considering $M^2\ll 1$, since we are dealing with microscopic $r_0$, we obtain expressions for $p(\tilde{r})$ and $\tilde{\rho}$ that allow (\ref{com}) to have the same  geodesics as the null geodesics of the Schwarzschild metric. We get 
\begin{align}
	\label{p}
	ds^2=-dt^2+\left(1-\frac{2M}{\tilde{\rho}}\right)^{-2}d\tilde{r}^2+\tilde{r}^2d\phi^2,
\end{align}
where
\begin{align}
	\tilde{\rho}=\frac{8\tilde{r}^3-54\tilde{r}^2M+2\tilde{r}-3M}{6}+\frac{6\left(\frac{\tilde{r}^2}{9}-\frac{\tilde{r}M}{2}\right)}{8\tilde{r}^3-54\tilde{r}^2M}\nonumber.
\end{align}
The last result has the following interpretation. For the Schwarzschild metric to be represented by the equation (\ref{discli3}), the  square root of the function associated with the radial coordinate on the equations (\ref{p}) and (\ref{discli3}) must be equal. Doing this equality and noticing that $q^{\frac{1}{\beta}}\ll1$ (for ordinary values of $\beta$ \cite{li}), we find the radial dependence of $q$ to be
\begin{align}	
\label{order}
\begin{split}
q(\tilde{r})&=\left[\frac{48M\tilde{r}^2\left(A+BT_c\right)}{\left(\Delta n\right)_0}\right]\times\left[32\tilde{r}^5-432M\tilde{r}^4+8\tilde{r}^3\right.\\
&-82M\tilde{r}^2\left.+2\tilde{r}-9M\right]^{-1},\\
%%%%&\left.+2\tilde{r}-9m\right]^{-1},
\end{split}
\end{align}
where again we considered $M^2 \ll 1$. This is the necessary scalar order parameter that must exist in a $(k=1,c=0)$ disclination-like configuration in order to the metric (\ref{discli3}) has  geodesics that coincides with the Schwarzschild null geodesics  for the consideration $M^2 \ll 1$, that is equivalent to the regime of weak-field approximation. Notice that, up to now, we have obtained the most general expression for the scalar order parameter $q$, since no consideration was made on what is creating this specific order parameter (flowing bulk, temperature and/or electric field gradient, etc.). Now, considering that we are dealing with a flowing liquid crystal, described by the case in the previous section, and substituting equation (\ref{order}) into (\ref{res}), we find the velocity profile that creates the order parameter (\ref{order}) to be
\begin{align}
	\begin{split}
\label{velox}	v(\tilde{r})&=\left\{\right.\left[144CM\tilde{r}^2\left(A+BT_c\right)\right]\times\left[\left(6\tilde{r}^5-1296M\tilde{r}^4\right.\right.\\
&\left.+24\tilde{r}^3-246M\tilde{r}^2+6\tilde{r}-27M\right)\left(\Delta n\right)_0\\
&-96M\tilde{r}^2\left(A+BT_c\right)\left.\right]^{-1}\left.\right\}^{\frac{1}{3\xi}}.
	\end{split}
\end{align}

However, when this bulk's velocity is close to light's velocity, drag effects must be considered \cite{novello}.

It is important to note that we could use the Gordon metric \cite{novello} to determine a velocity profile that simulates Schwarzschild's metric, since we are treating with a flowing medium and using the refractive index to calculate the behavior of light. However, the Gordon metric is used only when the refractive index of the medium is constant and here we are dealing with a varying refractive index, since \cite{kleman,satiro}
\begin{align}
\label{extra}
N_e^2(\vec{r})=n_o^2\cos^2\alpha+n_e^2\sin^2{\alpha},
\end{align}
where $\alpha$ is the angle between the propagation of light and the director $\vec{n}$, and $n_o$ and $n_e$ are no longer constant, because they depend on the scalar order parameter $q$.

Summarizing, every time we have a liquid crystal flowing toward a disclination, as described in the  previous section, and the bulk velocity is given by (\ref{velox}), any light ray traveling in this background will behave as in the presence of the Schwarzschild metric (\ref{schild}). But the key point is the equation (\ref{order}), that shows how the order parameter must be, on the conditions above cited, for the Schwarzschild metric to be simulated. 

%%%%%%%%%%%%%%%%%%%%%%%%%%%%%%%%%%%%%%%%%%%%%%%%%%%%%%%%%%%%%%%%%%%%
%%%%%%%%%%%%%%%%%%%%%%%%%%%%%%%%%%%%%%%%%%%%%%%%%%%%%%%%%%%%%%%%%%%%
\subsection{An Example}
%%%%%%%%%%%%%%%%%%%%%%%%%%%%%%%%%%%%%%%%%%%%%%%%%%%%%%%%%%%%%%%%%%%%
%%%%%%%%%%%%%%%%%%%%%%%%%%%%%%%%%%%%%%%%%%%%%%%%%%%%%%%%%%%%%%%%%%%%
Let's use the informations about the 5CB liquid crystal \cite{li,cang} and apply them in equation (\ref{velox}) to obtain the  velocity profile that simulates the Schwarzschild geometry. 

We consider that the liquid crystal system has a disclination-like configuration with a finite size core, but yet microscopic, and the liquid crystal flows inward this core. Using the possible values of order parameter when the temperature of the liquid crystal varies on the nematic phase \cite{kleman}, we admit that the maximum value of the order parameter $q(r)$ is $0.5$. Remembering this, we feed the equation (\ref{order}) with the 5CB data and, admitting that the maximum value for the order parameter is reached on the surface of the defect (in this example, the core measures $r_0=3\cdot10^{-5}\ m\Rightarrow\tilde{r}_0\approx5.12\cdot10^{-5}$), the equation (\ref{order}) provides us $M\approx1.13\cdot10^{-5}$. 

It's important to mention that $M$ is related to the Schwarzschild's radius, $r_{s}\equiv2M\approx2.26\cdot10^{-5}\ m<r_0=3\cdot10^{-5}\ m$, and how the light travels outside this radius, following an open path. Also, we notice that increasing $r_0$ or the value of the order parameter on the core's surface, makes $M$ increase.

\begin{figure}
	
		\includegraphics[width=0.7\textwidth]{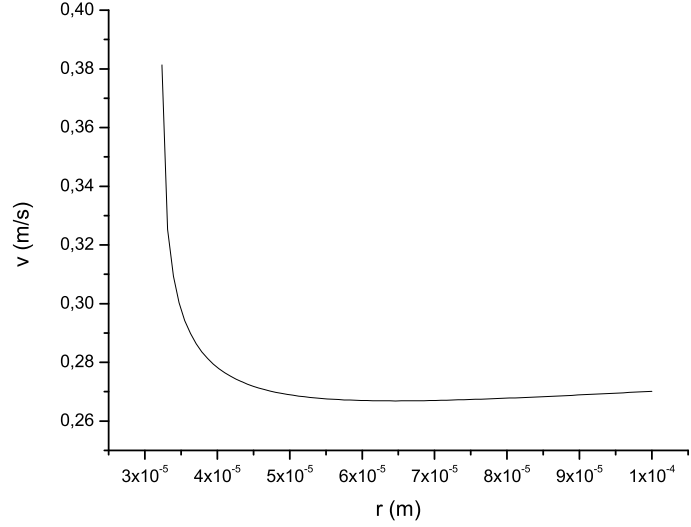}
		\caption{Velocity profile that simulates the Schwarzschild's metric around a disclination-like configuration with radius $r_0=3\cdot10^{-5}$ m.}
	\label{fig:1mps}
\end{figure}

Having $M$ on hands, the substitution of (\ref{order}) in (\ref{res}) gives $C$. However, this constant has an arbitrariness on its determination, because $C$ depends on the boundary conditions. Admitting that the bulk's velocity as the fluid reaches the drain to be $1\ m/s$, we get $C\approx0.334$. Finally, using all the constants found up to now in (\ref{velox}), we generate FIG. \ref{fig:1mps} which represents the velocity profile that simulates  Schwarzschild's metric. Because of the arbitrariness of $C$, any other value of this constant just shifts up or downward the points in this graph, holding the behavior of the velocity's gradients.

%%%%%%%%%%%%%%%%%%%%%%%%%%%%%%%%%%%%%%%%%%%%%%%%%%%%%%%%%%%

%%%%%%%%%%%%%%%%%%%%%%%%%%%%%%%%%%%%%%%%%%%%%%%%%%%%%%%%%%%

Now, let's evaluate the value of the fluid's velocity on the cylinder's surface that simulates the Schwarzschild metric obtained by Gordon metric. The algebraic expression for the flow velocity at a distance $r$ is \cite{rosquist}
\begin{align}
\label{gg}
	v_{Gordon}(r)=\frac{c}{\sqrt{1+\left(n^2-1\right)\frac{r}{2M}}},
\end{align}
where $n$ is the refractive index of the medium. The latter quantity should be $N_e$ since we are dealing with extraordinary light ray. Thus, equating $n$ to (\ref{extra}) for 5CB on the cylinder's surface and for an inward light ray ($\alpha=0$), we have $n=1.54$. Substituting this value, $M=1.13\cdot10^{-5}$ and $r=r_0=3\cdot10^{-5}$ m on (\ref{gg}), we obtain $v_{Gordon}=1.79\cdot10^{8}\ m/s$, a value $10^8$ times higher than one we obtained with $C=0.334$, $v=1\ m/s$. Therefore, the use of the Gordon metric in this problem is experimentally infeasible.

It's worth to note that the light in this flowing system describes open paths. Following the steps to calculate the light path as shown in most standard General Relativity books \cite{inverno,kip},  the values of $r_0$ and $M$ generate open trajectories. Thus, we are not concerned with investigating the existence of a horizon.

%%%%%%%%%%%%%%%%%%%%%%%%%%%%%%%%%%%%%%%%%%%%%%%%%%%%%%%%%%%%%%%%%%%%
%%%%%%%%%%%%%%%%%%%%%%%%%%%%%%%%%%%%%%%%%%%%%%%%%%%%%%%%%%%%%%%%%%%%
\section{Conclusion}
%%%%%%%%%%%%%%%%%%%%%%%%%%%%%%%%%%%%%%%%%%%%%%%%%%%%%%%%%%%%%%%%%%%%
%%%%%%%%%%%%%%%%%%%%%%%%%%%%%%%%%%%%%%%%%%%%%%%%%%%%%%%%%%%%%%%%%%%%
We have presented how a light beam in a special configuration of a flowing nematic liquid crystal can effectively behave as if in the presence of a Schwarzschild space-time. Using the equivalence of Fermat's principle and the existence of an effective metric \cite{satiro}, Haller's approximation allows us to use the Beris-Edwards' theory \cite{beris} to relate the properties of the effective metric to the local velocity of the liquid crystalline bulk. After that, Schwarzschild's metric described in terms of isotropic coordinates \cite{inverno,novello} permits us to use a cylindrically symmetric configuration of the nematic liquid crystal and to obtain the appropriate radial velocity profile around this configuration that gives us Schwarzschild's metric.

Remembering that we are dealing with velocities much lower than that of light, the distinguishing mark is the behavior of the bulk's velocity and not its absolute values. Thus, we could set velocity more appropriately within experimental possibilities.

As a complement of this research, the study of interactive forces between the configurations here analyzed without the restriction of a total topological charge (otherwise, it's known as the dipole-dipole attractive force \cite{poulin}) can be realized, as well as the theoretical study of some \emph{classical tests} of general relativity \cite{inverno} applied to this flowing system.

This work was partially supported by  CNPq, CAPES, REDE NANOBIOTEC BRASIL and INCT-FCx.


\begin{thebibliography}{0}
\bibitem{nota} An extraordinary light ray is the second kind of light ray that arises from the propagation in an anisotropic media - like a liquid crystal in its nematic phase - and it indicates the direction of the Poynting vector $\vec{S}$.
\end{thebibliography}
\end{document}